\documentclass[onecolumn,showpacs,amsfonts,aps,prc,nofootinbib,floatfix,%
superscriptaddress]{revtex4}

\usepackage{amsmath}
\usepackage{bm}
\usepackage{graphicx}

\usepackage{epsfig}
\newcommand{\beq}{\begin{equation}}
\newcommand{\eeq}{\end{equation}}
\newcommand{\bea}{\vspace{0.25cm}\begin{eqnarray}}
\newcommand{\eea}{\end{eqnarray}}

\newcommand{\ro}{\mbox{{\boldmath
$\rho$}}}

\newcommand{\kb}{{{\bf k}}}

\def\lsim{\mathrel{\rlap{\lower4pt\hbox{\hskip1pt$\sim$}}
    \raise1pt\hbox{$<$}}}         
\def\gsim{\mathrel{\rlap{\lower4pt\hbox{\hskip1pt$\sim$}}
    \raise1pt\hbox{$>$}}}         

\newcommand{\landau}{L.D.~Landau Institute for Theoretical Physics,
        GSP-1, 117940, Kosygina Str. 2, 117334 Moscow, Russia}

\begin{document}


\title{
Monte Carlo Glauber model with meson cloud:
predictions for 5.44 TeV Xe+Xe collisions
}
\date{\today}

\author{B.G.~Zakharov}\affiliation{\landau}

\begin{abstract}

We study, within the Monte-Carlo Glauber model, centrality dependence of the 
midrapidity charged  
multiplicity density $dN_{ch}/d\eta$  and of the anisotropy coefficients $\varepsilon_{2,3}$
in Pb+Pb collisions at $\sqrt{s}=5.02$ TeV
and in Xe+Xe collisions at $\sqrt{s}=5.44$ TeV.
Calculations are performed for
versions with and without nucleon meson cloud.
The fraction of the binary collisions, $\alpha$, has been fitted to the data 
on  $dN_{ch}/d\eta$ in Pb+Pb collisions.
We obtain $\alpha\approx 0.09(0.13)$ with (without) meson cloud.  
The effect of meson cloud on the $dN_{ch}/d\eta$
is relatively small. 
For Xe+Xe collisions for $0$-$5$\% centrality bin we obtain 
$dN_{ch}/d\eta\approx 1149$ and $1134$ 
with and without meson cloud, respectively. 
We obtain $\varepsilon_2(\mbox{Xe})/\varepsilon_2(\mbox{Pb})\sim 1.45$
for most central collisions, 
and   $\varepsilon_2(\mbox{Xe})/\varepsilon_2(\mbox{Pb})$
close to unity at $c\gsim 20$\%.
We find a noticeable increase of the eccentricity in Xe+Xe
collisions at small centralities due to the prolate shape
of the Xe nucleus. 
The triangularity in Xe+Xe collisions 
is bigger than in Pb+Pb collisions at $c\lsim 70$\%.
We obtain $\varepsilon_3(\mbox{Xe})/\varepsilon_3(\mbox{Pb})\sim 1.3$ at
$c\lsim 1$\%.

\end{abstract}
%

\maketitle

\section{Introduction}
It is believed that 
production of soft particles
in $AA$ collisions at RHIC and LHC energies occurs via formation
of the quark-gluon plasma (QGP) that  
expands hydrodynamically as a  near-ideal liquid \cite{Huov_hydro,Kodama_hydro}.
Hydrodynamic models have been successfully used for
description of the data from RHIC and LHC 
on centrality dependence
of hadron multiplicities and flow effects 
in $AA$ collisions.
The currently available LHC data on $AA$ collisions have
been obtained for Pb+Pb collisions at $\sqrt{s}=2.76$ and $5.02$ TeV.
Recently at the LHC there has been performed a run for Xe+Xe collisions 
($A=129$) at $\sqrt{s}=5.44$ TeV. The data from this run will allow to 
study variation of the $A$-dependence of soft hadron production 
in $AA$ collisions. 
One can expect that 
the flow effects for Xe+Xe collisions should be stronger 
than for Pb+Pb collisions because fluctuations in the 
initial entropy density should increase for nuclei with 
a smaller nucleon number.
For this reason the data on Xe+Xe collisions    
are of great interest for testing the hydrodynamic picture
of the QGP fireball evolution.
The Xe+Xe collisions at $\sqrt{s}=5.44$ TeV 
have been discussed recently in Refs. \cite{Olli_Xe,Eskola_Xe}.
 
The hydrodynamic simulations of $AA$ collisions require imposing
the initial conditions for the entropy/energy distribution at the QGP 
production time $\tau_0\sim 0.5-1$ fm  \cite{Heinz_hydro2,Heinz_tau}.
One of the popular approach for setting the initial conditions for the 
QGP fireball in $AA$ collisions is the Monte-Carlo Glauber (MCG) wounded 
nucleon model  
\cite{PHOBOS_MC,GLISS,GLISS2,Enterria_MC}.
In Refs.~\cite{MCGL1,MCGL2} we have developed
a version of the MCG wounded nucleon model for nucleons with meson cloud.
The meson-baryon Fock components of the nucleon play an important 
role in the flavor dependence of nucleon parton distribution functions (PDFs) 
in deep inelastic scattering \cite{ST}, and allow to explain 
the violation of the Gottfried sum rule \cite{ST}.
It is important
that, similarly to the wounded nucleon model
with constituent quarks
\cite{woun_q1,woun_q2,Voloshin_3q,Bozek_qD,Bozek_2016_3q,Loizides_q},
the meson degrees of freedom
lead to a nonlinear  increase of $dN_{ch}(AA)/d\eta$ 
with the number of the wounded nucleons \cite{MCGL2}.
This effect should emerge independently of the specific mechanism of 
inelastic interactions. It is important that, contrary to the MCG models with
the quark subnucleon degrees of freedom, the interaction
of the meson components is better understood, say, 
within the quark-gluon string model \cite{Kaidalov,Capella}.
Similarly to the ordinary two-component MCG model  without meson cloud 
\cite{GLISS2}
the model \cite{MCGL1,MCGL2} accounts for the contributions from soft
interaction (participant wounded nucleons) and from hard binary collisions
\cite{KN}.
However, the results of \cite{MCGL1,MCGL2} show that in the presence
of the meson-baryon Fock components the required fraction of the binary
collisions, $\alpha$, becomes smaller.
The results of \cite{MCGL2} show that
the meson cloud may improve somewhat agreement with
the data on the dependence of the elliptic flow on the charged multiplicity
for very small centralities defined via the ZDCs signals
for collisions of the deformed uranium nuclei at $\sqrt{s}=0.193$ TeV
\cite{UU_STAR}.
In the present paper we apply the model of Refs.~\cite{MCGL1,MCGL2} 
to obtain predictions
for Xe+Xe collisions at $\sqrt{s}=5.44$ TeV. 
To fix the parameters of the model we use the data on $pp$ 
and Pb+Pb collisions.

The plan of the paper is as follows. In Sec.~2 we 
outline the theoretical framework.
In Sec.~3 we present the numerical results.
We give conclusions in Sec.~4.

\section{Outline of the model}
In this section we briefly sketch our MCG scheme. 
We refer the reader to Ref.~\cite{MCGL2} for more details.

We represent the physical nucleon wave function $
|N_{phys}\rangle$
in the infinite-momentum frame (IFM) 
as the Fock-state composition of a bare nucleon $|N\rangle$
and an effective two-body meson-baryon $|MB\rangle$ state
\beq
|N_{phys}\rangle=\sqrt{1-n_{MB}}|N\rangle+
\int dxd\kb\Psi_{MB}(x,\kb)|MB\rangle\,,
\label{eq:10}
\eeq
where $x$ is the fractional longitudinal meson momentum 
in the physical nucleon, 
$\kb$ is the tranverse meson momentum, $\Psi_{MB}$ is
the IMF (light-cone) wave function of the $MB$ Fock state, and
\beq
n_{MB} =\int dxd\kb|\Psi_{MB}(x,\kb)|^2
\label{eq:20}
\eeq
is the total weight of the $MB$ Fock state. 
The previous analyses of 
the meson cloud effects in deep inelastic scattering (for a review,
see Ref.~\cite{ST}) 
show that the total weight of the meson-baryon Fock states 
in the nucleon is $\sim 40$\%. Due to this,  
we take for the probability of the effective 
$MB$ state $n_{MB}=0.4$.
Since the meson-baryon component is dominated by the 
pion-nucleon $\pi N$ state \cite{ST}, we 
calculate the  IMF distribution of the effective $MB$ state
for the $\gamma_5$ spin vertex. 
As in  the analyses of the deep inelastic scattering  \cite{ST}, we introduce  
a phenomenological vertex formfactor, $F$,
to account for the internal structure of the hadrons.
We take it in the form \cite{ST}
\beq
F=\left(\frac{\Lambda^2+m_N^2}{\Lambda^2+M^{2}_{\pi N}}\right)^2\,,
\label{eq:30}
\eeq
where $M_{\pi N}$ is the invariant mass of the $\pi N$ system.
We take $\Lambda=1.3$ GeV, supported by
the analysis \cite{HSS_MB} of the data on the process $pp\to nX$.
This value, at the same time, is also supported by the data on 
the violation of the Gottfried sum rule for the nucleon PDFs
\cite{ST}. 
Note, however, that our results are not very sensitive to the value 
of $\Lambda$.

In our model, as in the well known version of the 
MCG wounded nucleon model at the elementary nucleon-level 
GLISSANDO \cite{GLISS2}, 
the entropy is 
deposited in the soft sources 
from participants (related 
to soft interactions) and in the hard sources from the binary collisions, 
related to hard reactions, between the colliding particles. However, 
in our scheme the inelastic interaction of the physical nucleons
from the colliding objects  may occur as $N+N$, $N+MB$, $MB+N$ and
$MB+MB$ collisions. 
Because at high energy the midrapidity 
multiplicity density for all baryons and mesons should be similar,
we assume that the constituents $M$ and $B$ in the $MB$ effective 
Fock state interact like a pion and a nucleon, respectively.
We assume that the inelastic cross sections
for the bare baryon and meson states obey the constituent quark 
counting rule $4\sigma^{NN}_{in}=6\sigma^{MB}_{in}=9\sigma^{MM}_{in}$.
We use the Gaussian impact parameter  profile for the probability of $ab$ 
inelastic interaction of the bare constituents
\beq
P_{ab}(\rho)=\exp\left(-\pi \rho^2/\sigma_{in}^{ab}\right)\,.
\label{eq:40}
\eeq
The parameter $\sigma_{in}^{NN}$ has been 
adjusted to fit the experimental inelastic $pp$ cross section 
for non-single-diffractive (NSD) events.
The use of the data on the NSD $pp$ events
is reasonable because the diffractive events do not contribute to the 
multiplicity in the midrapidity region that we consider.

We assume isentropic evolution of the QGP fireball. Then 
the initial entropy rapidity density produced in an $AA$ collision is 
proportional to the final charged multiplicity pseudorapidity density
\beq
dS/dy=C dN_{ch}/d\eta\,,
\label{eq:50}
\eeq
where $C\approx 7.67$ \cite{BM-entropy}. 
For this reason in our calculations we consider the soft and the hard 
sources of the entropy as direct
sources of the multiplicity density. 
We will consider the charged multiplicity density 
$dN_{ch}/d\eta$ at the central pseudorapidity $\eta=0$
defined as $N_{ch}$ in the unit pseudorapidity 
window $|\eta|<0.5$.
We assume that the sources generated in all possible
collisions of the bare constituents (i.e., for $NN$, $MN$, and $MM$
collisions) have the same intensity.
This approximation is supported by the calculations 
within the quark-gluon string model \cite{Kaidalov,Capella} which show
that the difference between the midrapidity multiplicity density 
generated in $NN$, $MN$ and $MM$ interactions is small.

We model the fluctuations of the charged particle 
density generated by the sources by the Gamma distribution
\beq
\Gamma(n,\langle n\rangle)=
\left(\frac{n}{\langle n\rangle}\right)^{\kappa-1}
\frac{\kappa^\kappa\exp\left[-n\kappa/\langle n\rangle\right]}
{\langle n\rangle \Gamma(\kappa)}\,,
\label{eq:60}
\eeq
which is widely used in the MCG simulations.
For each soft source corresponding to  a wounded constituent 
the contribution to the multiplicity density is given by 
$(1-\alpha)\Gamma/2$, and for a hard source from a binary collision
it is simply $\Gamma$. However, for each pair of 
wounded particles the probability of a hard binary collision is 
suppressed by $\alpha$.
The parameters $\langle n\rangle$ and $\kappa$ 
have been adjusted 
to reproduce the experimental $pp$ data on the mean charged multiplicity 
and its variance in the unit pseudorapidity window $|\eta|<0.5$,
and the value of $\alpha$ has been fitted from the data on 
Pb+Pb collisions (see below).

For calculations of the multiplicity density $dN_{ch}/d\eta$
one can use the approximation of the point-like sources. 
But the smearing of the sources may be important
in calculations 
of the initial anisotropy coefficients $\varepsilon_n$ of the QGP fireball,
which in terms of the spacial entropy distribution
$\rho_s=dS/dyd\ro$ read
\cite{Teaney_en,Ollitraut_en}
\beq
\varepsilon_n=\frac{\left|\int d\ro \rho^n e^{in\phi}\rho_s(\ro)\right|}
{\int d\ro \rho^n\rho_s(\ro)}\,,
\label{eq:70}
\eeq
where the transverse vectors $\ro$ are calculated in the 
c.m. frame, i.e., $\int d\ro \ro\rho_s(\ro)=0$. 
To model the smearing of the sources we 
use a Gaussian source distribution
\beq  
\exp{\left(-\ro^2/\sigma^2\right)}/\pi \sigma^2\,.
\label{eq:80}
\eeq
We perform calculations for
$\sigma=0.7$ and $0.4$ fm.
The results for the anisotropy coefficients 
become sensitive 
to $\sigma$ only for very peripheral 
$AA$ collisions.

We perform calculations for the Woods-Saxon nuclear distributions with 
the hard-core repulsion.
Following to Ref.~\cite{GLISS2}, we take
for the hard-core radius $d=0.9$ fm.  
For the one-body Woods-Saxon distributions 
for $^{129}$Xe we use the $\theta$-dependent nuclear density 
\beq
\rho_{A}(r,\theta)=\frac{\rho_0}{1+\exp[(r-R_A(\theta)/a]}\,,
\label{eq:90}
\eeq
\beq
R_A(\theta)=R[1+\beta_2Y_{20}(\theta)+\beta_4Y_{40}(\theta)]
\label{eq:100}
\eeq
with $Y_{20}$ and $Y_{40}$ the spherical harmonics, 
with $\beta_2=0.162$, and $\beta_4=-0.003$ \cite{AtData}.
For the $^{208}$Pb nucleus we use the ordinary spherically symmetric
Woods-Saxon formula
with a $\theta$-independent ($\beta_{02}=\beta_{04}=0$) radius $R_A$. 
For the nucleus radii we use the formula 
$R=(1.1A^{1/3}-0.656/A^{1/3})$ fm, and take 
$a=0.459$ fm borrowed from Ref.~\cite{GLISS2}. To understand
the role of the prolate shape of the Xe nucleus, we also perform
 calculations for Xe+Xe collisions using the spherically symmetric
Woods-Saxon formula. 
As will be seen from our results the prolate shape of the Xe nucleus
increases noticeably the ellipticity $\varepsilon_2$ at small centralities.

\section{Results}
The direct $pp$ data on the charged multiplicity and inelastic cross
section
at 
$\sqrt{s}=5.02$ and $5.44$ TeV are absent. 
We obtained $dN_{ch}/d\eta$ for these energies with the help of the power 
law interpolation 
$dN_{ch}/d\eta \propto s^{\delta}$
between the ALICE data \cite{ALICE_nch541} at $\sqrt{s}=2.76$ TeV
($dN_{ch}/d\eta \approx 4.63$)  and 
at $\sqrt{s}=7$ TeV   
($dN_{ch}/d\eta=5.74\pm 0.15$) for the charged multiplicity in NSD events. 
It gives 
$dN_{ch}/d\eta
[\sqrt{s}=5.02\,,5.44\,\mbox{TeV}] \approx [5.32,\,5.42]
$.
We used a similar procedure to obtain the NSD $pp$ inelastic cross sections
at $\sqrt{s}=5.02$ and $5.44$ TeV from the ALICE \cite{ALICE4968}
results for $\sqrt{s}=2.76$ TeV ($\sigma_{in}^{NSD}\approx 50.24$ mb)  
and $7$ TeV ($\sigma_{in}^{NSD}\approx 58.56$ mb). We obtained the values:
$\sigma_{in}^{pp}[\sqrt{s}=5.02\,,5.44\,\mbox{TeV}] \approx [54.44,\,56.18]$ mb.

Since the difference in the energy between 
Xe+Xe collisions at $\sqrt{s}=5.44$ TeV and 
Pb+Pb collisions at $\sqrt{s}=5.02$ TeV is relatively small, we 
use the same fraction of the binary 
collisions $\alpha$ for $\sqrt{s}=5.02$ and $5.44$ TeV.
We determine $\alpha$ from fits to the ALICE \cite{ALICE_Pb5} data on 
the centrality dependence
of the midrapidity charged multiplicity density $N_{ch}$ at $|\eta|<0.5$
in Pb+Pb collisions $\sqrt{s}=5.02$ TeV.
As in Ref.~\cite{MCGL2}, we use a two step procedure.
First, we fitted $\langle n\rangle$ and $\kappa$ 
for a broad set of $\alpha$
to the $N_{ch}$ in $pp$ collisions  imposing the condition $N_{ch}/D=1$,
which is well satisfied for $|\eta|<0.5$ window 
\cite{UA1_pp,ALICE_nch541}.
Then, we used the values of $\langle n\rangle$ and $\kappa$ to fit 
the parameter $\alpha$ from the Pb+Pb data.
This procedure gives 
$\alpha\approx0.09$ and $\alpha\approx 0.13$  for the scenarios 
with and without meson cloud, respectively.
The parameters of the Gamma distribution
(\ref{eq:60}) obtained from the fit with meson 
cloud to the $pp$ data for the above optimal value $\alpha=0.09$ 
read: $\langle n\rangle\approx 4.69(4.8)$, 
$\kappa\approx 0.516(0.524)$ for $\sqrt{s}=5.02(5.44)$ TeV.
For the scenario without meson cloud for the optimal value
$\alpha=0.13$ we obtained
$\kappa\approx 0.56(0.56)$ for $\sqrt{s}=5.02(5.44)$ TeV, 
(in the version without meson cloud the value of $\langle n\rangle$ 
is simply equal to the experimental $N_{ch}$ for $pp$ collisions).

In Fig.~1 we show the results
of our fit to the ALICE data on the centrality dependence
of the midrapidity charged multiplicity density in 
Pb+Pb collisions at $\sqrt{s}=5.02$ TeV \cite{ALICE_Pb5}
obtained by Monte Carlo generation 
of $\sim 2\cdot 10^6$ events for the scenarios with and without meson cloud.
The scenario 
with meson cloud gives somewhat better 
agreement with the data ($\chi^2/\mbox{d.p}\approx 0.1$)
as compared to the version without meson cloud
($\chi^2/\mbox{d.p}\approx 0.3$).
\begin{figure}
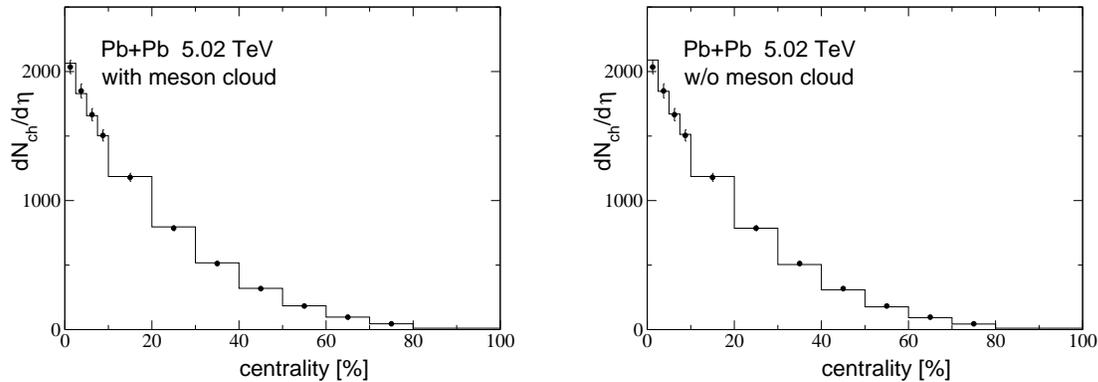

\hspace*{-0.8cm }\epsfig{file=fig1a.eps,height=5cm,clip=,angle=0} 
\hspace*{0.8cm } \epsfig{file=fig1b.eps,height=5cm,clip=,angle=0} 
\caption{\small Centrality dependence of midrapidity $dN_{ch}/d\eta$ for 
Pb+Pb collisions at $\sqrt{s}=5.02$ TeV. Left: 
MCG simulation for the scenario with meson cloud
at $\alpha=0.09$.
Right: MCG simulation for the scenario without meson cloud
at $\alpha=0.13$. Data are from ALICE \cite{ALICE_Pb5}.
}
\end{figure}
\begin{figure}
\epsfig{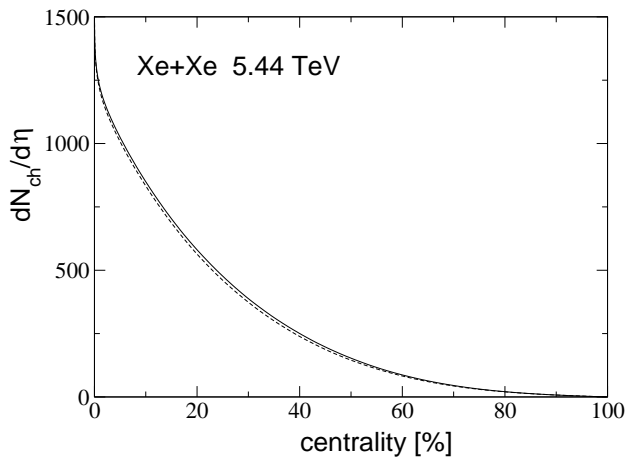}
\caption{\small Centrality dependence of 
 midrapidity $dN_{ch}/d\eta$ for Xe+Xe collisions at $\sqrt{s}=5.44$ TeV
obtained 
for the scenarios with (solid) meson cloud at $\alpha=0.09$ 
and without (dashed) meson cloud at $\alpha=0.13$.
}
\end{figure}
\begin{figure}[ht]
\hspace*{-0.8cm }\epsfig{file=fig3a.eps,height=5cm,clip=,angle=0} 
\hspace*{0.8cm } \epsfig{file=fig3b.eps,height=5cm,clip=,angle=0} 
\caption{\small Centrality dependence of the rms $\varepsilon_2$ for
the Gaussian source distribution (\ref{eq:80}) for $\sigma=0.7$ fm
for Pb+Pb collisions at $\sqrt{s}=5.02$ TeV (solid)
and Xe+Xe collisions at $\sqrt{s}=5.44$ TeV 
for the $\theta$-dependent (dashed) and symmetric Woods-Saxon
distribution (dotted).
Left: the version with meson cloud 
at $\alpha=0.09$. Right: the version without meson cloud at $\alpha=0.13$.
}
\end{figure}
\begin{figure}[ht]
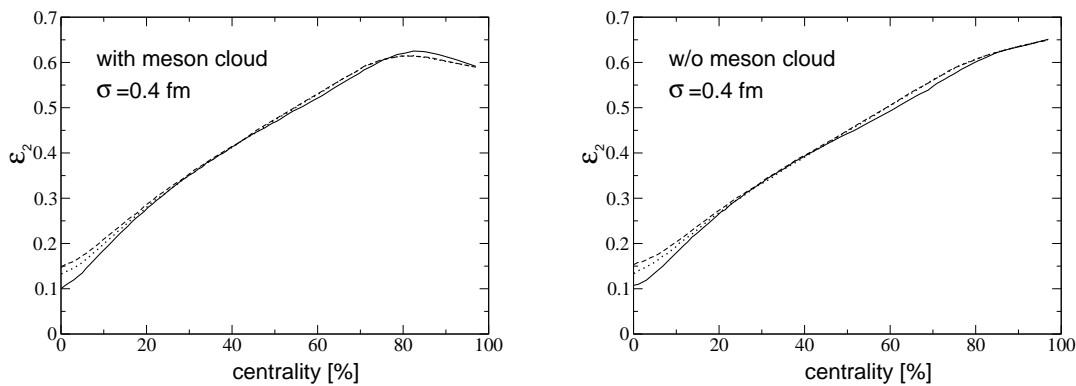

\hspace*{-0.8cm }\epsfig{file=fig4a.eps,height=5cm,clip=,angle=0} 
\hspace*{0.8cm } \epsfig{file=fig4b.eps,height=5cm,clip=,angle=0} 
\caption{\small Same as in Fig.~3 but for $\sigma=0.4$ fm.
}
\end{figure}
\begin{figure}[ht]
\hspace*{-0.8cm }\epsfig{file=fig5a.eps,height=5cm,clip=,angle=0} 
\hspace*{0.8cm } \epsfig{file=fig5b.eps,height=5cm,clip=,angle=0} 
\caption{\small Centrality dependence of the rms $\varepsilon_3$ for
the Gaussian source distribution (\ref{eq:80}) for $\sigma=0.7$ fm
for Pb+Pb collisions at $\sqrt{s}=5.02$ TeV (solid)
and Xe+Xe collisions at $\sqrt{s}=5.44$ TeV 
for the $\theta$-dependent Woods-Saxon
distribution (dashed). Left: the version with meson cloud 
at $\alpha=0.09$. Right: the version without meson cloud at $\alpha=0.13$.
}
\end{figure}
\begin{figure}[ht]
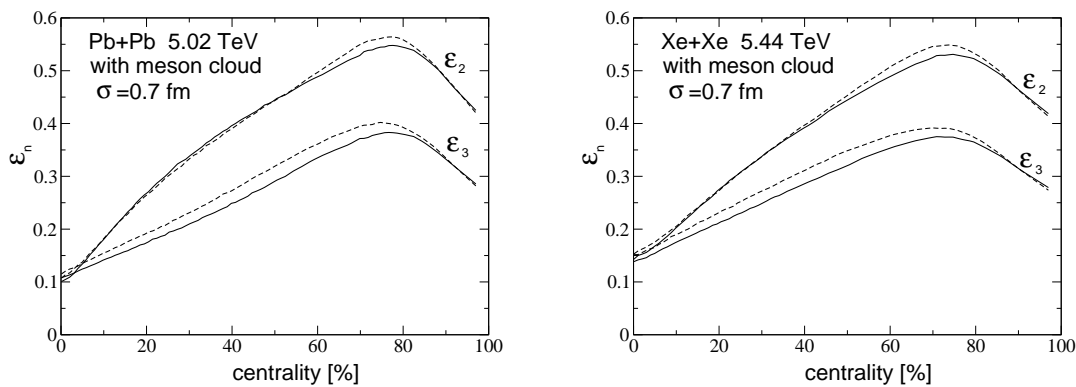

\hspace*{-0.8cm }\epsfig{file=fig6a.eps,height=5cm,clip=,angle=0} 
\hspace*{0.8cm } \epsfig{file=fig6b.eps,height=5cm,clip=,angle=0} 
\caption{\small 
Centrality dependence of the rms $\varepsilon_{2,3}$ 
for the version with meson cloud
with the Gaussian source distribution (\ref{eq:80}) for $\sigma=0.7$ fm
in Pb+Pb collisions at $\sqrt{s}=5.02$ TeV (left)
and in Xe+Xe collisions at $\sqrt{s}=5.44$ TeV (right)
for the $\theta$-dependent Woods-Saxon
distribution with (solid) and without (dashed) the hard-core repulsion.
}
\end{figure}
\begin{figure}[ht]
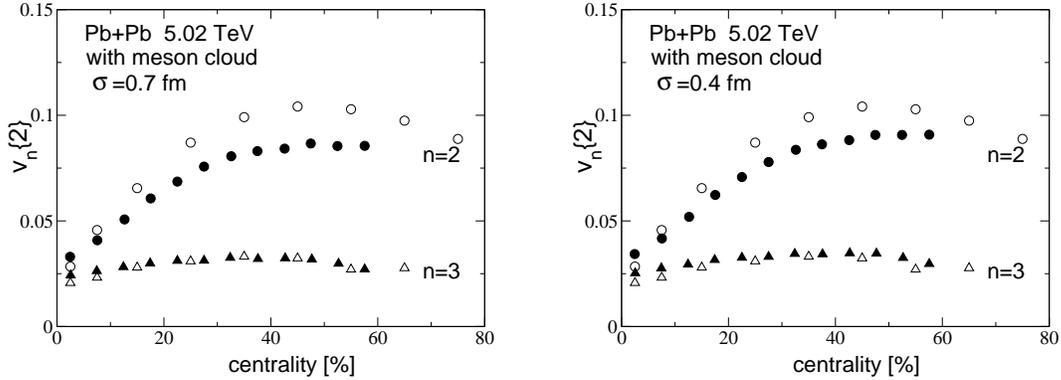

\hspace*{-0.8cm }\epsfig{file=fig7a.eps,height=5cm,clip=,angle=0} 
\hspace*{0.8cm } \epsfig{file=fig7b.eps,height=5cm,clip=,angle=0} 
\caption{\small Centrality dependence of 
$v_{2}\{2\}$ (filled circles) and $v_3\{2\}$ (filled triangles) 
for Pb+Pb collisions at $\sqrt{s}=5.02$ TeV
obtained from $\varepsilon_{2,3}$ with meson cloud
for $\sigma=0.7$ (left) and $0.4$ fm (right)
with the help of the linear response approximation 
(see text for explanations) using the results of the hydrodynamic
simulations of Ref.~\cite{Olli_Xe}.
Open circles and triangles
are $v_{2,3}\{2\}$ from ALICE \cite{ALICE_v2_5}.
}
\end{figure}
\begin{figure}[ht]
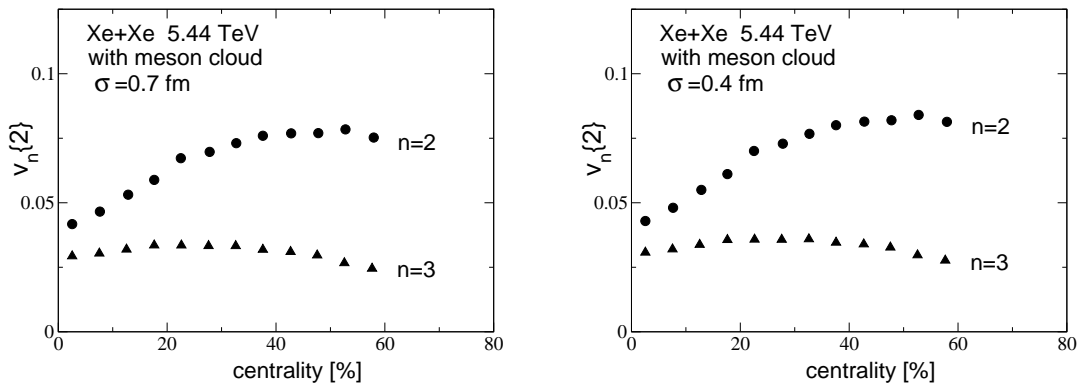

\hspace*{-0.8cm }\epsfig{file=fig8a.eps,height=5cm,clip=,angle=0} 
\hspace*{0.8cm } \epsfig{file=fig8b.eps,height=5cm,clip=,angle=0} 
\caption{\small Same as in Fig.~7 but for 
Xe+Xe collisions at $\sqrt{s}=5.44$ TeV.
}
\end{figure}

In Fig. 2 we show our predictions for centrality dependence of the
charged multiplicity in Xe+Xe collisions 
at $\sqrt{s}=5.44$ TeV obtained with and without meson cloud.
The difference between two version is relatively small. For
intermediate centrality region the meson cloud increases 
$dN_{ch}/d\eta$ by $\sim 5$\%. For the $0$-$5$\% centrality
$dN_{ch}/d\eta\approx 1149$ and $1134$ for the versions
with and without meson cloud, respectively. It is smaller by $\sim 6-7$\%
than the charged multiplicity density obtained in Ref.~\cite{Eskola_Xe}.
From Figs.~1,~2 one sees that as compared to Pb+Pb 
collisions at $\sqrt{s}=5.02$ TeV 
for Xe+Xe $0$-$5$\% central collisions
at $\sqrt{s}=5.44$ $dN_{ch}/d\eta$    becomes smaller by a factor of $\sim 1.7$. That 
corresponds to decrease of the initial QGP temperature by a factor of 
$\sim 1.2$. A remark is in order here. The curves in Fig.~2 are obtained
under assumption of an isentropic flow. But we fitted the parameters
of the model from the data on Pb+Pb collisions also ignoring the
non-isentropic effects. Since these effects are weak, possible errors
in the extrapolation of the results from Pb+Pb to Xe+Xe collisions
should be small.

In Figs. 3,~4 we present the results for 
the rms $\varepsilon_2$ 
(the rms $\varepsilon_n$ often denoted $\varepsilon_n\{2\}$, for clarity,
we omit $\{2\}$)  versus centrality for
Pb+Pb at $\sqrt{s}=5.02$ TeV and Xe+Xe at $\sqrt{s}=5.44$ TeV
for two version of the model. 
The anisotropy coefficients depend on the smearing parameter
$\sigma$ in (\ref{eq:80}).
We present the results for two values of  
the Gaussian width of the sources $\sigma=0.7$ and $0.4$ fm.
From Figs.~3,~4 one sees that for small centralities the results with 
and without
meson cloud are close to each other. For intermediate centralities 
the version with meson cloud
gives a little smaller $\varepsilon_2$. For very peripheral collisions
with centrality $\gsim 80$\% the model without meson
cloud gives bigger $\varepsilon_2$. 
From comparing the results in Figs.~3 and 4 one can see that 
the value of the smearing width becomes important at $c\gsim 30-40$\%, where
the eccentricity grows with decreasing $\sigma$.  
From Figs.~3,~4 one sees that for the most central collisions $c\lsim 5$\% 
$\varepsilon_s(\mbox{Xe})/\varepsilon_2(\mbox{Pb})\sim 1.3-1.4$, and 
$\varepsilon_s(\mbox{Xe})/\varepsilon_2(\mbox{Pb})$ becomes close to unity at $c\gsim 30$\%.
For very peripheral collisions with $c\gsim 75$\% 
$\varepsilon_2(\mbox{Xe})$ becomes a little smaller than $\varepsilon_2(\mbox{Pb})$.
From Figs.~3,~4, by comparing the curves for Xe+Xe collisions obtained 
for the asymmetric and symmetric Woods-Saxon distribution, 
one can see that the effect of the prolate shape of the Xe nucleus
on the eccentricity becomes important at $c\lsim 20$\%. For most central
collisions the asymmetric Woods-Saxon distribution increases the eccentricity
by $\sim 12$\%.
This is due to the highly overlapping body-body collisions  
of the nucleus ellipsoids, that lead naturally to an asymmetric entropy
deposition. 

In Fig.~5 we present the results for the rms triangularity $\varepsilon_3$
for $\sigma=0.7$ fm. One can see that both for Pb+Pb and Xe+Xe collisions
the triangularity for the versions
with and without meson cloud are very similar for small and intermediate
centralities. For very peripheral collisions 
$c\gsim 80$\% $\varepsilon_3$ for the version without meson cloud becomes
bigger. Our calculations show that for Xe+Xe collisions the effect
of the prolate form of the Xe nucleus is very small. For this reason
we do not plot in Fig.~5 the results for Xe+Xe collisions for the
symmetric Woods-Saxon distribution.
From Fig.~5 one sees that $\varepsilon_3(\mbox{Xe})/\varepsilon_3(\mbox{Pb})\sim 1.3$ for
most central collisions, and decreases to unity at $c\sim 70$\%.

The results shown in Figs.~1--5 have been obtained for the Monte-Carlo sampling
with the nuclear distribution with the hard-core radius $d=0.9$ fm.
It is evident that the presence of the hard-core repulsion should reduce
the fluctuations of the nuclear matter density in the colliding
nuclei, that can suppress the effect of the fluctuations in the initial
entropy deposition. To study the role of the hard-core repulsion we also
performed the calculations for $d=0$. In this case we also use the parameters
of the Woods-Saxon distribution from the analysis \cite{GLISS2}:
$R=(1.12A^{1/3}-0.86/A^{1/3})$ fm, and $a=0.54$ fm.  
Calculations in this version show that the effect of the hard-core repulsion
on the centrality dependence of the charged multiplicity density is small.
But the effect is stronger for the anisotropy coefficients. It is illustrated 
in Fig.~6 where we plot $\varepsilon_{2,3}$ versus centrality 
obtained with (solid) and without (dashed) the hard-core repulsion for 
the version with meson cloud at $\sigma=0.7$ fm.  
From  Fig.~6 one sees that the effect of the hard-core is bigger 
for the purely fluctuation-driven quantity $\varepsilon_3$.
At $c\lsim 70$\% the hard-core reduces $\varepsilon_3$ by $\sim 10$\%
both for Pb+Pb and Xe+Xe collisions.
For the eccentricity $\varepsilon_2$ the effect of the hard-core is noticeable
for small centralities. In this region $\varepsilon_2$ is also 
dominated by fluctuations, and  the hard-core repulsion  
reduces $\varepsilon_2$
by $\sim 10$\% both for Pb+Pb and Xe+Xe collisions.

Our results for the initial $\varepsilon_{2,3}$ for 
Pb+Pb and Xe+Xe collisions do not differ 
strongly from that of Refs.~\cite{Eskola_Xe,Olli_Xe}. 
The analysis \cite{Olli_Xe} is based on the TRENTO Monte-Carlo model 
\cite{TRENTO1,TRENTO2}, and \cite{Eskola_Xe} is based on the so 
called EKRT \cite{EKRT} model with mini-jet parton production.
For intermediate
centralities $c\sim 50$\% our $\varepsilon_2$ for $\sigma=0.7$ fm   
is smaller than that of Refs.~\cite{Eskola_Xe,Olli_Xe} by $\sim 10-15$\%.
But for small centralities our predictions for $\varepsilon_{2,3}$
are somewhat bigger.
Note that
in Refs.~\cite{Eskola_Xe,Olli_Xe} the calculations for Xe+Xe collisions 
have been performed  only 
with the symmetric Woods-Saxon nuclear distribution. For this reason
the grows of the ratio $\varepsilon_2(\mbox{Xe})/\varepsilon_2(\mbox{Pb})$ 
at $c\to 0$ in Refs.~\cite{Eskola_Xe,Olli_Xe} is not so strong as in our results.

The event by event hydrodynamic modeling of $AA$ collisions
shows that, except for very peripheral collisions, to a good approximation  
$v_n\approx k_n\varepsilon_n$ for $n=2,~3$ 
\cite{Heinz1_v2e2,Heinz2_v2e2,Niemi_v2e2,Ollitrault_v2e2}
(for a comprehensive review, see, e.g., Ref.~\cite{Yan}).
This linear response approximation works better for $v_2$ 
\cite{Niemi_v2e2,Ollitrault_v2e2}.
In the present paper we do not perform the hydrodynamic simulation
of Pb+Pb and Xe+Xe collisions that are necessary for accurate calculations
of $v_n$. To obtain predictions for the flow coefficients via 
our MCG results for $\varepsilon_n$ we use the above linear response relation
with $k_{2,3}$ defined via the ratio of $v_{2,3}\{2\}$ to rms $\varepsilon_{2,3}$
obtained 
for Pb+Pb collisions at $\sqrt{s}=5.02$ TeV and
for Xe+Xe collisions at $\sqrt{s}=5.44$ TeV 
in the hydrodynamic calculations
with the shear viscosity over entropy ratio $\eta/s=0.047$
in Ref.~\cite{Olli_Xe}
\footnote{In principle, it is clear
that this procedure should be quite accurate even beyond the linear response
picture (say, when the cubic terms become important \cite{Ollitrault_v2e2})
because the difference between our predictions for $\varepsilon_{2,3}$
and that of Ref.~\cite{Olli_Xe} is not very large.}. 
In Fig.~7 we compare $v_{2,3}\{2\}$ obtained in this way
for Pb+Pb collisions at $\sqrt{s}=5.02$ TeV 
with the results from ALICE \cite{ALICE_v2_5}. We used 
our anisotropy coefficients
$\varepsilon_{2,3}$ obtained with meson cloud for the Woods-Saxon
distribution with the hard-core repulsion for the smearing
width $\sigma=0.7$ and $0.4$ fm.
From Fig.~7 one sees that for $n=2$ the theoretical results underestimate
the data by $\sim 10-20$\% at $c\sim 30-60$\%, but for $n=3$ the agreement
is quite reasonable. 
The theoretical predictions for $v_{2,3}\{2\}$ in 
Xe+Xe collisions at $\sqrt{s}=5.44$ TeV
are shown in Fig.~8. The relation between our predictions for $v_n\{2\}$
to that of Ref.~\cite{Olli_Xe} similar to the situation with predictions 
for $\varepsilon_n$. For most central collisions, due to bigger values of 
$\varepsilon_{2,3}$, we predict somewhat bigger $v_{2,3}\{2\}$ (by $\sim 5$\% for
$\sigma=0.7$ fm and by $\sim 10$\% for $\sigma=0.4$ fm).

The predictions in Fig.~7,~8 correspond to the version with the hard-core 
repulsions. From the curves in Fig.~6 one sees that the hard-core
repulsion reduces the initial anisotropy coefficients due some reduction
of the the nuclear density  fluctuations. However, by no means 
the sampling of the nucleon positions with the Woods-Saxon
density with the hard-core repulsion can not be viewed as an accurate
method to account for the long range 
fluctuations in the colliding nuclei. For this reason
the predictions for the initial anisotropy coefficients 
may be questioned (especially for the fluctuation driven quantity 
$\varepsilon_3$).
Indeed, it is well known \cite{Greiner,Speth} that the dynamical long range 
fluctuations of the
nuclear matter in heavy nuclei are dominated by the giant
resonances, e.g., by the giant dipole, monopole, 
and quadrupole resonances. 
However these dynamical long range effects are completely ignored
in the Woods-Saxon distributions in the MCG model calculations.
In Ref.~\cite{Z_B} we have demonstrated that for the dipole mode 
of the $^{208}$Pb nucleus
the classical treatment based on the Monte-Carlo simulation with the Woods-Saxon
nuclear density overestimates the fluctuations
of the dipole moment squared by a factor of $\sim 5$.
The situation with the dynamical quantum effects for the 
monopole and quadrupole collective modes, that potentially may also 
be important, remains unclear.
It would be of great interest to study the role
of the dynamical quantum effects 
due to the giant resonances on the fluctuations
of the entropy deposition in $AA$ collisions
(of course, these effects may also be important for calculations
in other approaches, say, in the EKRT  model \cite{EKRT},
TRENTO model \cite{TRENTO2}, or in the color glass 
condensate scheme \cite{IP-GL1,IP-GL2}).
We leave this for future work.

\section{Conclusions}

We have studied the centrality dependence of the charged midrapidity 
multiplicity density in Pb+Pb collisions at $\sqrt{s}=5.02$ TeV
and in Xe+Xe collisions at $\sqrt{s}=5.44$ TeV within
the MCG model with and without meson cloud developed in Ref.~\cite{MCGL2}.
The parameters of the model have been fixed to the ALICE data 
\cite{ALICE_Pb5} on  $dN_{ch}/d\eta$ in Pb+Pb collisions.
We obtained the fraction of the binary
collisions $\alpha\approx0.09(0.13)$ with (without) meson cloud.  
With these parameters we give predictions for future LHC data
on Xe+Xe collisions at $\sqrt{s}=5.44$ TeV. 
We find that the effect of the meson cloud on the $dN_{ch}/d\eta$
is relatively small. For Xe+Xe collisions the meson cloud
increases $dN_{ch}/d\eta$ by $\sim 5$\% 
in the intermediate centrality region.
For the $0$-$5$\% centrality bin we obtained 
$dN_{ch}/d\eta\approx 1149$ and $1134$ 
with and without meson cloud, respectively. 
As compared to Pb+Pb 
collisions at $\sqrt{s}=5.02$ TeV 
for Xe+Xe $0$-$5$\% central collisions
at $\sqrt{s}=5.44$ $dN_{ch}/d\eta$   becomes smaller by a factor of $\sim 1.7$. 
It corresponds to decrease of the initial QGP temperature by a factor of 
$\sim 1.2$.

Both for Pb+Pb and Xe+Xe collisions
we do not find a significant effect of the meson cloud
on the $\varepsilon_{2,3}$ at $c\lsim 70$\%. But the meson cloud reduces
$\varepsilon_{2,3}$ for very peripheral collisions.
We find that the ratio of the eccentricity in Xe+Xe collisions
to that for Pb+Pb collisions is close to unity at $c\gsim 20$\%,
but it becomes bigger than unity at $c\lsim 20$\%.
We obtained $\varepsilon_2(\mbox{Xe})/\varepsilon_2(\mbox{Pb})\sim 1.45$
for most central collisions ($c\lsim 1$\%). 
We predict a noticeable increase of the eccentricity in Xe+Xe
collisions at small centralities due to the prolate shape
of the Xe nucleus. This effect gives $\sim 50$\% to
the difference between the eccentricity in most central 
Xe+Xe and Pb+Pb collisions.

We find that at $c\lsim 70$\%   the triangularity in Xe+Xe collisions 
is bigger than in Pb+Pb collisions. We obtain 
$\varepsilon_3(\mbox{Xe})/\varepsilon_3(\mbox{Pb})\sim 1.3$ at
$c\lsim 1$\% and $\varepsilon_3(\mbox{Xe})/\varepsilon_3(\mbox{Pb})\sim 1.1$
at $c\sim 50$\%.
We have investigated the effect of the hard-core repulsion in the Monte-Carlo
sampling of the nuclear distributions. We found that the hard-core repulsion
gives a relatively small effect on the charged multiplicity density.
But its effect is sizeable for the anisotropy 
coefficients $\varepsilon_{2,3}$.
For $\varepsilon_2$ the effect of the hard-core is noticeable
for small centralities where $\varepsilon_2$ is 
dominated by fluctuations. For most central collisions
the hard-core repulsion  
reduces $\varepsilon_2$ by $\sim 10$\% both for Pb+Pb and Xe+Xe collisions.
The triangularity $\varepsilon_3$ is reduced by 
the hard-core repulsion by
$\sim 10$\% at $c\lsim 70$\%  both for for Pb+Pb and Xe+Xe collisions.

To obtain predictions for the flow coefficients via 
our MCG results for $\varepsilon_n$ we have used the linear response relation
$v_n\approx k_{n}\varepsilon_n$ with $k_n$ defined via the ratio of 
$v_{2,3}\{2\}$ to rms $\varepsilon_{2,3}$
obtained in the recent hydrodynamic analysis \cite{Olli_Xe}.
The results for $v_{2,3}\{2\}$ obtained in this way
for Pb+Pb collisions at $\sqrt{s}=5.02$ TeV 
are in reasonable agreement with the data from ALICE \cite{ALICE_v2_5}.

\end{document}